\documentclass[10pt]{article}
\usepackage{epsfig}

\newcommand{\pt}{p_{\rm t}}
\def\beq{\begin{equation}}
\def\eeq#1{\label{#1}\end{equation}}
\def\eeqn{\end{equation}}

\def\beqa{\begin{eqnarray}}
\def\eeqa#1{\label{#1}\end{eqnarray}}
\def\eeqan{\end{eqnarray}}

\let\bar=\overbar

\def\Dslash{\not{\hbox{\kern-4pt $D$}}}
\def\dslash{\not{\hbox{\kern-2pt $\del$}}}

\def\msb{{\bar{\ssstyle M \kern -1pt S}}}

\def\Title#1{\begin{center} {\Large {\bf #1} } \end{center}}

\begin{document}

\Title{$\rm{D^{0}}$ cross section in pp collisions at $\sqrt{s}=7$ TeV, measured with the ALICE experiment}

\bigskip\bigskip


\begin{raggedright}

{\it Xianbao Yuan, for the ALICE Collaboration\index{Yuan, X.}\\
Institute of Particle Physics, Central China Normal University \\
University and INFN, Padova, Italy \\
Contact:yuanxb@iopp.ccnu.edu.cn or xianbao.yuan@pd.infn.it}
\bigskip\bigskip
\end{raggedright}

The measurement of the cross-section for charm production in pp
collisions at the LHC is not only a fundamental reference to investigate
medium properties in heavy-ion collisions, but also a key test of pQCD
predictions in a new energy domain.

The ALICE~\cite{ch507} experiment has measured the D meson production in pp collisions at $\sqrt{s}=7$ TeV.
We present the analysis procedure for $\rm D^{0}\to \rm K^{-}\rm \pi^{+}$
and for the calculation of efficiency and acceptance corrections. Finally, we show the
preliminary results on $\rm D^0$ cross section in pp collisions at $\sqrt{s}=7$ TeV,
measured in the region $2<p_{\rm t}<12$ GeV/$c$ at central rapidity $|y|<0.5$. These results are
compared to perturbative QCD predictions.

The analysis is based on an invariant mass analysis of
opposite-charge pairs of reconstructed tracks that can
represent a $\rm D^0$ with a displaced vertex (the mean proper decay length of the $\rm D^0$ is
$c\tau\approx 123~\mu$m). The selection is based on topological cuts and particle identification
via specific energy deposit and time-of-flight measurement. The cross
section is calculated from the raw signal yield extracted with the
invariant mass analysis, $N^{\rm D^{0}~raw}(p_{\rm t})$, using the following formula:
\begin{equation}
  \left.\frac{{\rm d}\sigma^{\rm D^0}}{{\rm d}\pt}\right|_{|y|<0.5}=
  \frac{1}{2}\frac{1}{2\, y_{\rm acc}\Delta\pt}\frac{\left.f_{prompt}\cdot N^{\rm D^{0}~raw}(\pt)\right|_{|y|<y_{\rm acc}}}{\epsilon_{prompt}\cdot{\rm BR} \cdot\mathit{L}_{int}}\cdot
\end{equation}

Here, $\epsilon_{prompt}$ means the efficiency of prompt mesons, which accounts for selection cuts, for track and primary vertex
reconstruction efficiency, and for detector acceptance. The $f_{prompt}$ is the prompt fraction of raw yield.

Figure~1 (Left) shows the invariant mass distribution for $p_{\rm t}>2$ GeV/$c$ after applying the cuts,
which corresponds to $1.1\times 10^8$ minimum bias events collected by ALICE in 2010 at $\sqrt{s}=7$ TeV.
Figure~1 (Right) shows the efficiencies for $\rm D^{0}\to \rm K^{-}\rm \pi^{+}$ with all the decay particles in the
acceptance $\left|\eta\right|<0.9$. The efficiencies increase and flatten at about 0.1 at $p_{\rm t}>2$ GeV/$c$.
The efficiency without particle identification selection, shown for comparison, is the
same as with particle identification for $p_{\rm t}>2$ GeV/$c$, indicating that this selection
is essentially fully efficient for the signal. The efficiencies for $\rm D^0$ meson from B meson
decay, also shown for comparison, are larger by a factor about 2, because this feed-down
component is more displaced from the primary vertex, due to the longer B life time. The $10-15\%$ feed-down from
B decays is subtracted based on pQCD prediction~\cite{ch508}.

Several sources of systematic uncertainties were considered, namely those affecting
the signal extraction from the invariant mass spectra and all the correction factors applied
to obtain the $p_{\rm t}$-differential cross sections. A summary of the estimated relative systematic
errors is shown in Fig~2 (Left).

The $p_{\rm t}$-differential cross section for prompt $\rm D^0$, obtained from the yields extracted
by fitting the invariant mass spectra and corrected for efficiency and B feed-down,
is shown in Fig~2 (Right). The error bars represent the statistical errors, while the systematic
errors are plotted as rectangle areas around the data points. The measured $\rm D^{0}$ meson production cross section is compared to two theoretical predictions, namely FONLL~\cite{ch508} and GM-VFNS~\cite{ch513}. Our measurement of $\rm D^0$ at
$\sqrt{s}=7$ TeV is reproduced by both models within their theoretical uncertainties.

\begin{figure}[!t]
\begin{center}
    \leavevmode
    \includegraphics[height=5cm,width=0.4\textwidth]{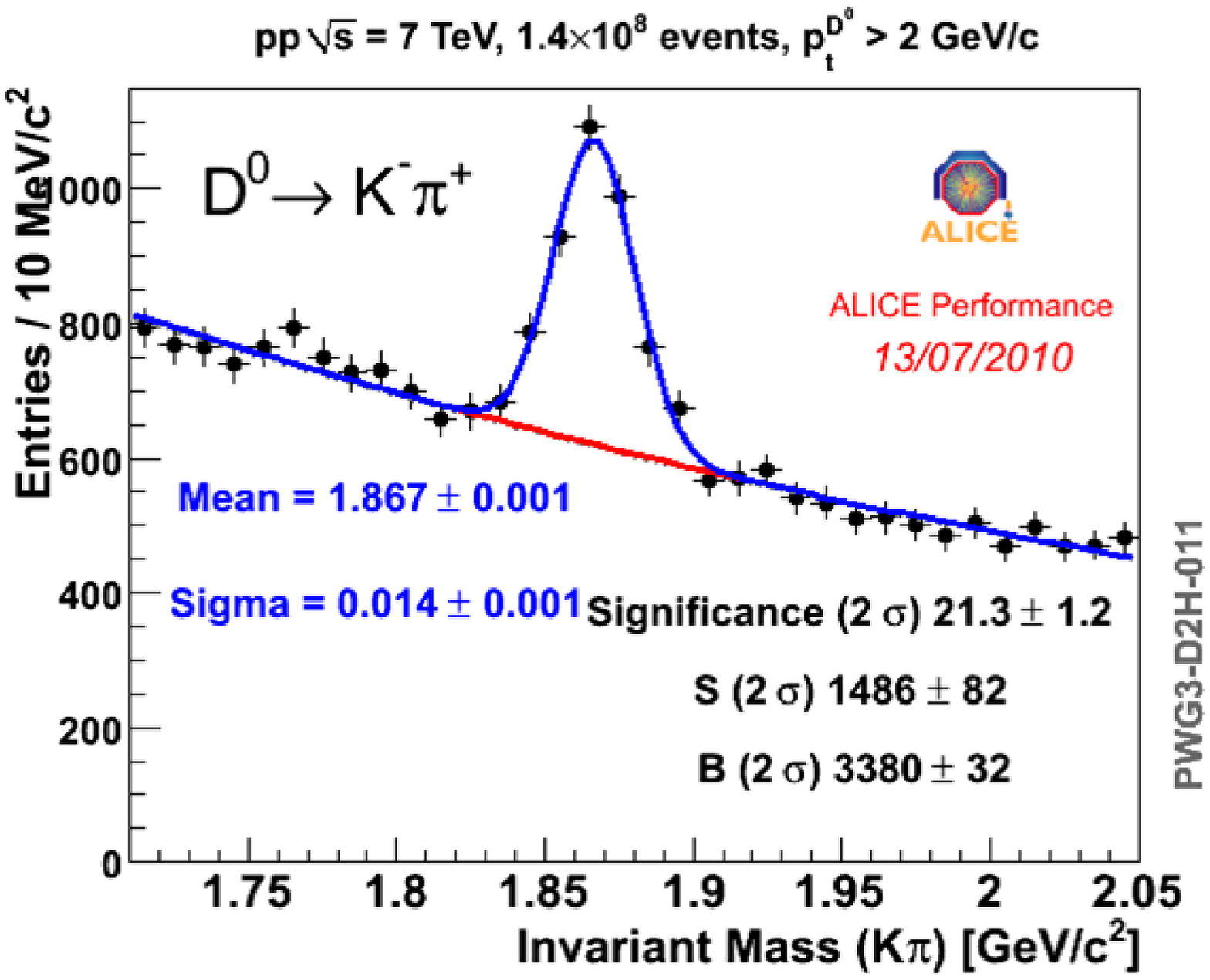}
    \hspace{0.005\textwidth}
    \includegraphics[height=5cm,width=0.4\textwidth]{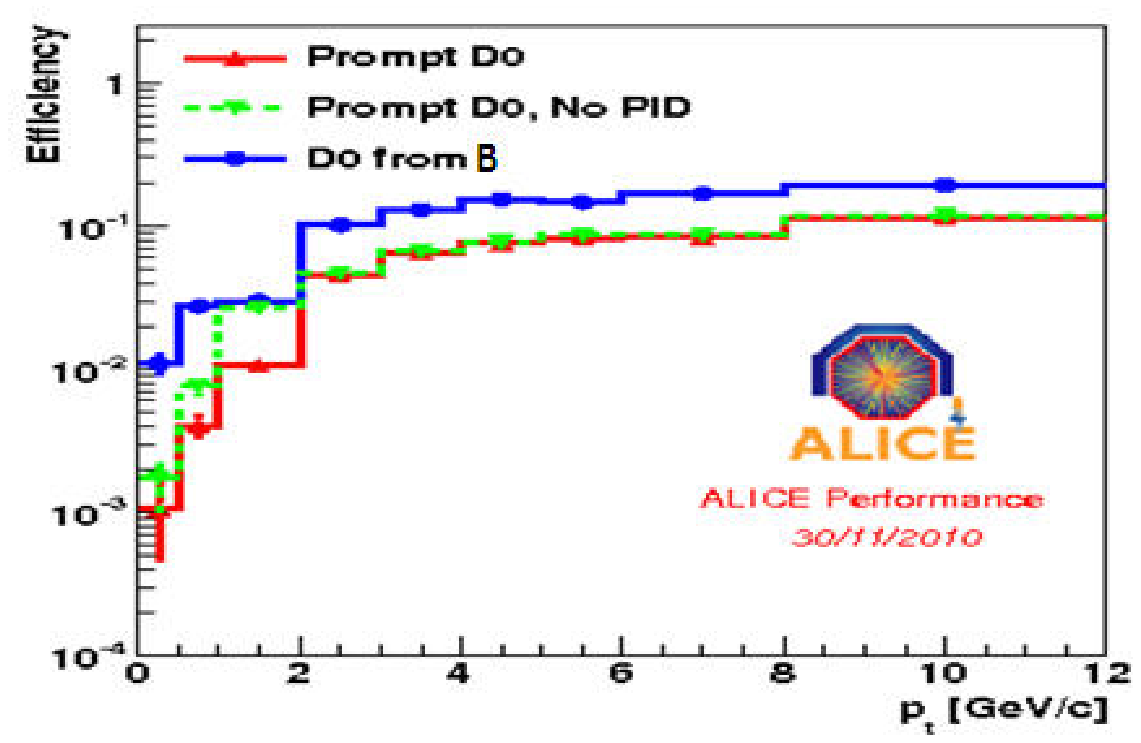}
   \caption{Left: $\pt>$2 GeV/$c$ invariant mass distribution. Right: efficiencies for $\rm D^0$ as a function of $p_{\rm t}$.}
    \label{fig:d0Gaussians}
  \end{center}
\end{figure}

\begin{figure}[!t]
 \begin{center}
   \leavevmode
   \includegraphics[height=5cm,width=0.45\textwidth]{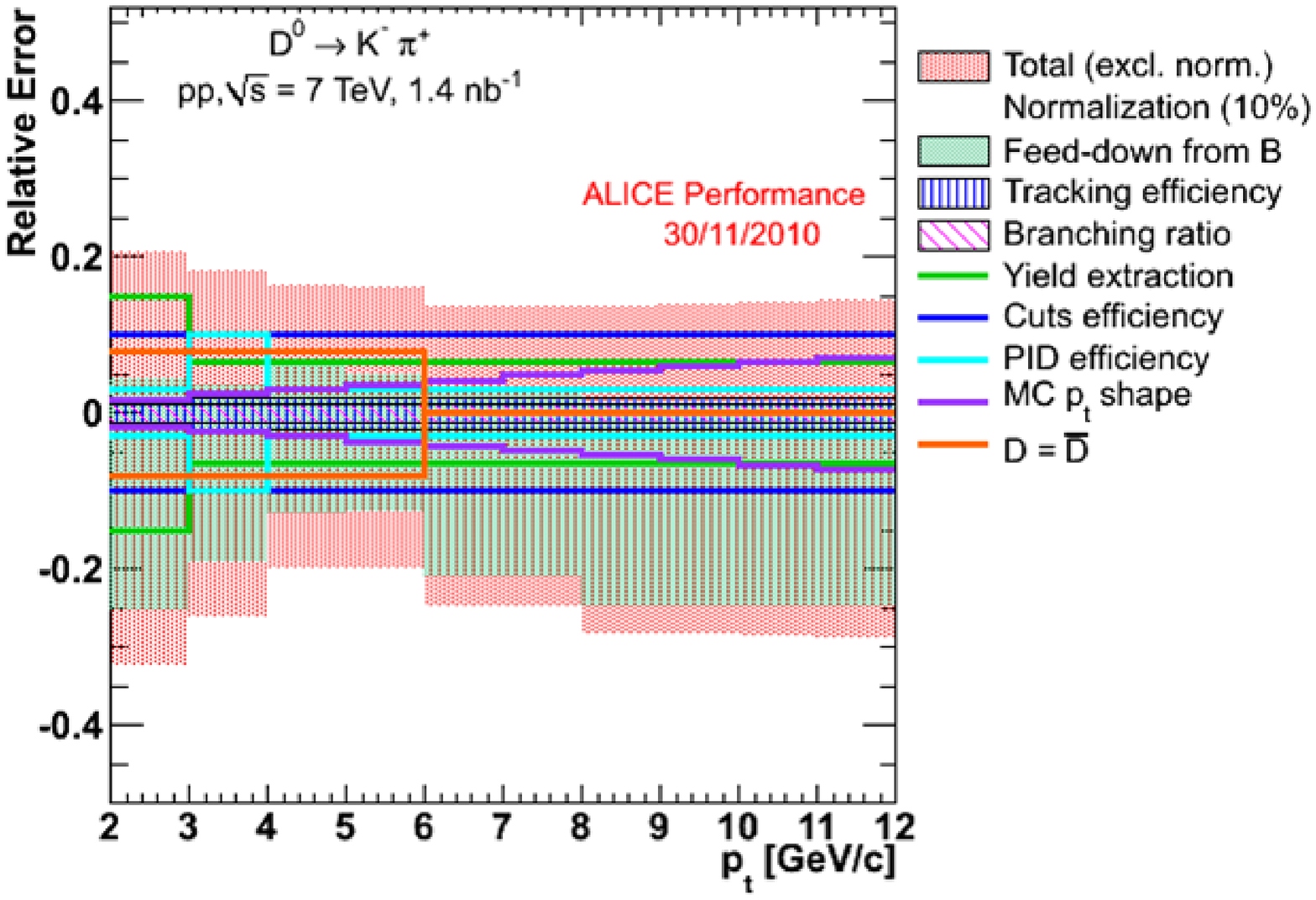}
    \hspace{0.005\textwidth}
    \includegraphics[height=5cm,width=0.4\textwidth]{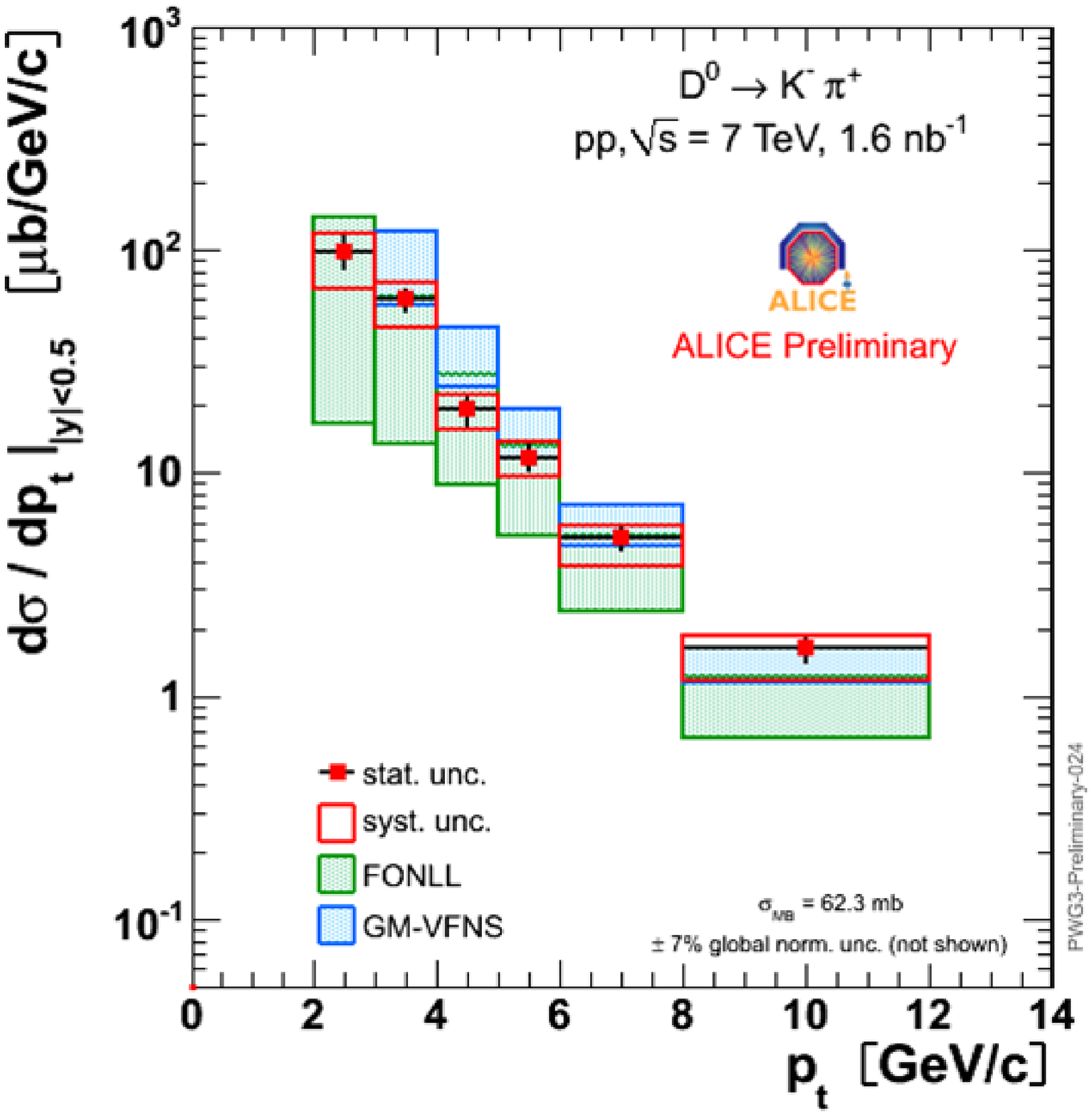}
   \caption{Left: systematic errors summary plot. Right: $p_{\rm t}$-differential cross section for prompt $\rm D^0$
   in pp collisions at $\sqrt{s}=7$ TeV compared with FONLL~\cite{ch508} and GM-VFNS~\cite{ch513} theoretical predictions.}
    \label{fig:d0Gaussians}
  \end{center}
\end{figure}

\end{document}